\documentclass[aps,pre,showpacs]{revtex4}
\usepackage{epsfig}
\usepackage{amsmath,amssymb,amsthm}
\usepackage{graphicx}
\usepackage{psfrag}
\usepackage{bm}
\def\n{\langle n \rangle}
\def\nop{\langle n_{OP} \rangle}
\def\s2{\sigma^2}

\begin{document}
\title{Critical wetting of a class of nonequilibrium interfaces: a mean-field picture}

\author{Francisco de los Santos$^{1,2}$, Elvira Romera$^{2,3}$, Omar Al Hammal$^{1,2}$, 
and Miguel \'Angel Mu\~noz$^{1,2}$ }

\affiliation{$^1$Departamento de Electromagnetismo y F{\'\i}sica de la
Materia, Universidad de Granada, Fuentenueva s/n, 18071 Granada,
Spain \\
$^2$Instituto Carlos I de F{\'\i}sica Te\'orica y
Computacional, Universidad de Granada, Fuentenueva s/n, 18071 Granada,
Spain  \\
$^3$ Departamento de F\'isica At\'omica, Molecular y Nuclear, 
Universidad de Granada, Fuentenueva s/n, 18071 Granada, Spain}

\date{\today}

\begin{abstract}

A self-consistent mean-field method is used to study critical wetting
transitions under nonequilibrium conditions by analyzing
Kardar-Parisi-Zhang (KPZ) interfaces in the presence of a bounding
substrate. In the case of positive KPZ nonlinearity a single
(Gaussian) regime is found. On the contrary, interfaces corresponding
to negative nonlinearities lead to three different regimes of critical
behavior for the surface order-parameter: (i) a trivial Gaussian
regime, (ii) a weak-fluctuation regime with a trivially located
critical point and nontrivial exponents, and (iii) a highly
non-trivial strong-fluctuation regime, for which we provide a full
solution by finding the zeros of parabolic-cylinder functions. These
analytical results are also verified by solving numerically the
self-consistent equation in each case. Analogies with and differences from 
equilibrium critical wetting as well as nonequilibrium complete
wetting are also discussed.

\end{abstract}
\pacs{02.50.Ey,05.50.+q,64.60.-i}
\maketitle

\section{INTRODUCTION}

Wetting transitions can be considered as an interface unbinding transition
where the mean interfacial separation from a substrate plays the role of an order-parameter.
In the wet state the interface drifts away from the substrate, while in 
the nonwet state the interface is bound to it, with many contact points
between them. In equilibrium, such phenomena can be described by Langevin-type 
equations of the form \cite{reviews_wetting}

\begin{equation}
\frac{\partial h (x,t)}{\partial t}= D \nabla^2 h -\frac{\delta V(h)}{\delta h} + \eta(x,t)
\label{ew_wall}
\end{equation}
with $V(h)$ standing for the energy between the substrate and the
interface, $\eta$ being a Gaussian white noise, and $D$ a diffusion
coefficient. Alternatively, the number of contact points between the
substrate and the interface can also be regarded as an order-parameter
that equals zero when $\langle h \rangle=\infty$ and is nonzero
otherwise.  This quantity is closely related to the surface
order-parameters studied in the framework of lattice systems in a
semi-infinite geometry, for which the interfacial displacement models
like Eq. (\ref{ew_wall}) are expected to be useful effective
descriptions \cite{binder}.  A familiar example is provided by Ising
ferromagnets where the surface order-parameter is the average
magnetization at the substrate.  The variable $\n = \langle e^{-h}
\rangle$ constitutes an adequate mathematical representation of such
an order-parameter that
exhibits singular behavior: it is a positive
quantity that vanishes as $\n \sim |a-a_c|^\beta$ close to criticality,
with $a$ denoting a convenient control parameter, 
and evolves in time as $\n \sim t^{-\theta}$ right at the transition.

An extension to nonequilibrium situations was considered by adding the
nonlinear term $\lambda (\nabla h)^2$ to Eq. (\ref{ew_wall}),
which represents preferential growth along the local normal to the
surface and covers the realm of Kardar-Parisi-Zhang (KPZ) interfacial
phenomena \cite{kpz,varios}. This is expected to capture the physics of
wetting transitions under nonequilibrium circumstances in the simplest
possible form. Interestingly, in this case a simple Langevin equation
for $n$ does exist if short-ranged forces between the substrate and
the interface are assumed. One typically writes
\cite{reviews_wetting},
\begin{equation}
V(h) = \int dx \left[a h(x)+be^{-ph(x)}\right]
\end{equation}
where $a$ and $b$ are phenomenological
parameters. The change of variables $h={\rm sgn}(\lambda) \ln n$ leads to an equation 
of the form
(see \cite{reviews} for details), 

\begin{equation}
\frac{\partial n(x,t)}{\partial t} =D\nabla^2 n +an +bn^{1+p}
+n\eta(x,t),
\label{mn}
\end{equation}
to be interpreted in the Stratonovich sense. Here, $b$ is a
temperature dependent-parameter, $a$ plays the role of a chemical
potential difference and is the control parameter, and $\langle
\eta(x,t)\rangle=0$, $\langle \eta(x,t) \eta(x',t')\rangle =\sigma^2
\delta(x-x')\delta(t-t')$.  $p$ carries the opposite sign of $\lambda$
which is the ultimate responsible for the physical behavior. For
$p>0$, the above equation represents the so called {\it multiplicative
noise 1} (MN1) universality class \cite{varios,reviews}. For $p<0$,
$n$ is merely an auxiliary variable that diverges at the transition
and it is $1/n$ which has to be studied instead. In this case the
associated set of exponents define the {\it multiplicative noise 2}
(MN2) universality class \cite{fattah,alemanes,mn2}. A detailed
discussion of the differences between the MN1 and MN2 universality
classes can be found in \cite{reviews}. In the following discussion
the surface order-parameter will be denoted $n_{OP}$.  It should be
emphasized that $n_{OP}=n$ for MN1 and $n_{OP}= 1/n$ for MN2.  Despite
the simple appearance of Eq. (\ref{mn}), we nonetheless caution
that the intricacies and subtleties of the KPZ equation lie in the
interplay between the diffusion and the multiplicative Gaussian white
noise terms.

The MN1 and MN2 universality classes are examples of nonequilibrium,
complete wetting transitions, in which $a$ has to be fine
tuned. A second type of wetting transitions may occur if as the
temperature is increased $b$ becomes bigger and eventually changes its
sign while the system is kept at coexistence, $a=a_c$. This is
denoted {\em critical wetting} and amounts to taking $b<0$ and adding a
higher-order term $cn^{1+2p}$ in order to obtain a finite solution for
$n$. Equivalently, the potential to be considered in the interfacial
representation is $V(h)=ah+be^{-ph}+ce^{-2ph}$, where the last term
ensures stability. More precisely, the model system is defined by the
Langevin equation

\begin{equation}
\frac{\partial n(x,t)}{\partial t} =D\nabla^2 n +an +bn^{1+p}+cn^{1+2p}
+n\eta(x,t),
\label{critical_mn}
\end{equation}
in which now $b$ is the new control parameter and $a$ is set to the
critical value $a_c$ found for the complete wetting transition. For
sufficiently low values of $b<0$ the interface remains pinned and the
density of locally pinned sites at the wall is high. As the transition
is approached (increasing $b$), the stationary density of pinned
segments goes to zero in a continuous manner as $\langle
n_{OP}(b,t=\infty) \rangle \sim |b-b_c|^\beta$. Above $b_c$, the
interface depins and therefore the mean separation $\langle h \rangle$
diverges and $\nop $ vanishes. In this latter case the density of
pinned segments scales with time as $\langle n_{OP}(b>b_c,t) \rangle
\sim t^{-\theta}$, with the exponent $\theta$ adopting different
values for MN1 ($p>0$) and MN2 ($p<0$).

In this paper we study within the mean-field approximation the critical
wetting transition associated with Eq. (\ref{critical_mn}). In
the absence of exact solutions, mean-field approaches are useful 
not only in enabling analytic calculations to be performed, but also
because they provide insight into the physical behavior at high system
dimensionalities which would be otherwise unattainable from computer
simulations alone. Since, in the present case, it is known that
mean-field theory could be valid for dimensions as low as $d=2$ ($d=3$
bulk dimensions) at least in some regime, the results presented here
may be relevant for realistic three-dimensional systems
\cite{reviews}.

This paper is organized as follows. In Sec. II we describe the
mean-field approach and provide details of the calculation for both
the MN1 and MN2 cases. As will be proved, three different scaling
regimes have to be distinguished for MN1, and only one for MN2.
Results for higher moments of $\nop$ and for $\langle h \rangle$ are
also included. Section III contains a summary and a discussion of our
findings.


\section{MEAN-FIELD APPROACH}

Sound mean-field approximations to multiplicative-noise equations like (\ref{critical_mn}) 
require that the effects of both the noise and of the spatially varying order-parameter
are taken into account to some extent. To this effect, the following procedure can be used \cite{vdbroeck}:
the Laplacian is discretized as $1/2d \sum_j(n_j-n_i)$, where
$n_i=n(x_i,t)$ and the sum is over the nearest-neighbors of $i$. Afterward, the value of 
the nearest-neighbor is substituted by the average field $\langle n \rangle$
to obtain a closed Fokker-Planck equation for $P(n,t,\langle n \rangle)$. 
The steady-state solution is then found from the self-consistency requirement \cite{birner}

\begin{equation}
\langle n \rangle = \frac{\int_0^\infty n P(n,\n )}
{\int_0^\infty  P(n,\n)}.
\label{self-cons}
\end{equation}
In what follows we particularize Eq. (\ref{self-cons}) to the MN1
and MN2 cases [see Eq. (\ref{critical_mn})]. It will be shown that for
MN1 there is a sequence of scaling regimes depending on the relative
importance of the noise strength as compared to the spatial coupling
$D$ and the nonlinearity exponent $p$: (i) a pure mean-field regime
where the noise can be completely disregarded, (ii) a weak-noise
regime where the noise strength enters the expression of the
exponents, but without shifting the wetting temperature, and (iii) a
strong noise regime where both the wetting temperature and the
exponents are noise dependent.  For MN2 the situation is far less
rich, exhibiting a single mean-field-like scaling regime.

In the following analysis the mean-value theorem for infinite integrals \cite{gradshteyn}
will be used repeatedly to determine the asymptotic behavior in $\n$ of the 
various integrals. According to this theorem, under quite general integrability and 
boundedness conditions,

\begin{equation}
\int_a^\infty dx \ f(x)g(x)=\mu \int_a^\infty dx \ g(x),
\end{equation}
where $\mu$ is some value between the lower and upper bounds of $f(x)$ \cite{gradshteyn}.
Likewise, the combination $2D/\s2$ will be denoted by $\nu$ to simplify the notation.

\subsection{The case MN1} 

For the MN1 case ($p>0$), the associated stationary probability density can be 
readily obtained from the associated associated Fokker-Planck equation \cite{vkp} and reads

\begin{equation}
P_{st}(n) \propto n^{\gamma-1} \exp
\bigg\{-\frac{2}{\s2}\bigg( \frac{b}{p} n^p+\frac{c}{2p}n^{2p}+\frac{Dm}{n}
\bigg)\bigg\},
\end{equation}
where $m=\n$, and $\gamma = -2(a+D)/\sigma^2$.
After defining 

\begin{equation}
I(m) =\int_0^\infty dn \  n^\gamma \exp\bigg\{
-\frac{2b}{\sigma^2 p} n^p-\frac{c}{\sigma^2p}n^{2p}-\frac{2D}{\sigma^2}\frac{m}{n}
\bigg\},
\end{equation}
and substituting $n=\nu mx$, $I(m)=(\nu m)^{1+\gamma} J(m)$, with

\begin{equation}
J=\int_0^\infty dx \ x^\gamma e^{-b'(mx)^p-c'(mx)^{2p}}e^{-1/x}, 
\quad b'=\frac{2b}{\s2 p} \nu^p, 
\quad c'=\frac{c}{\s2 p} \nu^{2p}.
\label{j}
\end{equation}
The self-consistency equation (\ref{self-cons})
can now be recast in the simpler form

\begin{equation}
-\frac{\nu}{m}=\frac{1+\gamma}{m} + \frac{\partial_m J(m)}{J(m)}.
\label{consistency}
\end{equation}

Since the mean-field, self-consistent calculation for complete wetting (MN1) \cite{birner}
yields $a_c=\sigma^2/2$ ($ \gamma_c+1=\nu$) and $J>0$ does not diverge, the condition (\ref{consistency}) 
simplifies to $\partial_m J=0$. 

We now consider an intermediate point $x_1>0$ such that $mx_1 \ll 1$ and split
$J(m)$ as 

\begin{equation}
J=J_1+J_2=\int_0^{x_1}dx \ x^{-1-\nu} e^{-b'(mx)^p-c'(mx)^{2p}}e^{-1/x}
+ \int_{x_1}^\infty dx \ x^{-1-\nu} e^{-b'(mx)^p-c'(mx)^{2p}}e^{-1/x},
\label{split}
\end{equation}
after which $\exp[-b'(mx)^p-c'(mx)^{2p}]$ in $J_1$ is expanded to second order, whereupon 

\begin{equation}
J_1\simeq
c_1-b'c_2m^p-c_3m^{2p} +O(m^{3p}),
\end{equation}
where $c_1, c_2$, and $c_3$ are constants. 
As for $J_2$, since the derivative can enter the integral,
\begin{eqnarray}
\frac{dJ_2}{dm} &=& \int_{x_1}^\infty dx \ x^{-1-\nu} e^{-b'(mx)^p-c'(mx)^{2p}} e^{-1/x} 
\Big( -p b' m^{p-1}x^p-2pc'm^{2p-1}x^{2p}\Big)  
\nonumber \\
&=& e^{-1/\xi(m)} \int_{x_1}^\infty dx \ x^{-1-\nu}  e^{-b'(mx)^p-c'(mx)^{2p}} 
\Big( -pb'm^{p-1}x^p-2pc'm^{2p-1}x^{2p} \Big),
\end{eqnarray}
with $\xi(m) \in [x_1,\infty)$ being a by-product of the application of
the mean-value theorem. Taking $(mx)^p=t$,

\begin{equation}
\frac{dJ_2}{dm}= -\frac{e^{-1/\xi(m)}}{m^{1-\nu}} \int_{(mx_1)^p}^\infty dt \ t^{-\frac{\nu}{p}-1} e^{-b't-c't^2} 
\Big(b't+2c't^2\Big) = \frac{c_4(m)}{m^{1-\nu}}.
\label{dj2dm}
\end{equation}
Finally,
\begin{equation}
\frac{d J}{dm}= pb'c_2m^{p-1}+2pc_3m^{2p-1}+c_4(m) m^{\nu-1} = 0.
\label{complete}
\end{equation}
To proceed further requires identifying the term involving the lowest power of $m$,
which in turn requires studying how $c_4(m)$ modifies $m^{\nu-1}$ for small values of $m$. 

First, notice that the factor $\exp(-1/\xi)$ is innocuous in such a limit.
Second, we work out the low-$m$ limit of the two integrals contained in $c_4(m)$, 
namely,

\begin{equation}
c_{\nu}= \int_{(mx_1)^p}^\infty dt \ t^{-\frac{\nu}{p}}  \ e^{-b't-c't^2}
, \qquad
c_{\nu+1}= \int_{(mx_1)^p}^\infty dt \ t^{-\frac{\nu}{p}+1} \ e^{-b't-c't^2}, 
\end{equation}
to find after splitting,

\begin{equation}
\int_{(mx_1)^p}^\infty dt \ t^\delta e^{-b't-c't^2}= 
\int_{(mx_1)^p}^1 dt \ t^\delta  e^{-b't-c't^2}  +\int_1^\infty dt \ t^\delta
e^{-b't-c't^2}.
\label{j2sim}
\end{equation}
After applying the mean-value theorem to the first integral a contribution $m^{p(1+\delta)}$ 
is obtained (the second integral contributes a constant),
whence it ensues that the leading asymptotic behavior of Eq. (\ref{complete}) 
is unaffected by $c_4(m)$ and hence two cases must be distinguished.

{\em Case 1:} $p <\nu$ leads to an equation  of the form ($\bar{c}_2$,$\bar{c}_3$, and $\bar{c}_4$
being positive factors)
\begin{equation}
b\bar{c}_2+m^p \bar{c}_3 + m^{\nu-p}\bar{c}_4=0, 
\end{equation}
which implies $m\sim (-b)^\beta$ as $b\to 0$, with
\begin{equation}
\beta =
\begin{cases}
\frac{1}{p} & {\rm if} \quad p < \frac{D}{\s2} \cr
\frac{1}{\nu-p} & {\rm if} \quad \frac{D}{\s2} < p < \frac{2D}{\s2} \cr
\end{cases}
\end{equation}

{\it Case 2:} $p>\nu$. It is expedient to rewrite Eq. (\ref{dj2dm}) as
 
\begin{eqnarray}
\frac{dJ_2}{dm}&=&
-\frac{e^{-1/\xi(m)}}{m^{1-\nu}} 
\int_{(mx_1)^p}^\infty dt \ t^{(1+\gamma)/p-1} e^{-b't-c't^2} \Big( b't+2c't^2\Big)= 
\frac{-e^{-1/\xi}}{m^{1-\nu}} \nonumber \\
&& \times \bigg\{ 
\int_0^\infty dt \ t^{-\nu/p-1} e^{-b't-c't^2} \Big(b't+2c't^2\Big)
-\int_0^{(mx_1)^p} dt \ t^{-\nu/p-1} e^{-b't-c't^2} \Big(b't+2c't^2\Big) 
\bigg\}. \nonumber \\
\end{eqnarray}
The first integral, which we call $c_5(b)$, does not depend on $m$ but can vanish for particular
values of $b$, while the for the second one we again apply the mean-value theorem
to obtain the leading (lowest) powers $m^{p-\nu}$, $m^{2p-\nu}$ when $m\sim 0$.
Last,

\begin{equation}
b \tilde{c}_2 m^{p-\nu}+\tilde{c}_3  m^{2p-\nu}+c_5(b)=0.
\end{equation}
Note that the factors $\tilde{c}_2$ and $\tilde{c}_3 $ are given in terms of exponentials of  $\xi_1 \in [x_1,\infty)$,$\xi_2 \in [0,(mx_1)^p]$ which result from the application of the mean-value theorem, 
and do not vanish as $m\to 0$. 
The next step is to find out what values of $b$ make $c_5(b)$ vanish. The latter
consists of two integrals that can be easily written in terms of 
parabolic-cylinder functions $D_\mu(x)$, using the recurrence relation
$D_\mu(x)=xD_{\mu-1}(x)+(1-\mu)D_{\mu-2}(x)$ \cite{gradshteyn}:

\begin{equation}
c_5(b)=(2c')^{\frac{\nu}{2p}} \ \Gamma\bigg(1-\frac{\nu}{ p}\bigg) \
e^{\frac{b^2}{2 c \sigma^2 p}}
D_{\frac{\nu}{p}}\Bigg(b \sqrt{\frac{2}{ c\sigma^2 p}}\Bigg) .
\end{equation}
Therefore, the critical point is determined by the zeros of the parabolic-cylinder 
functions. It turns out that $D_{\nu/p}$ has exactly one zero of order one on 
the interval at hand \cite{zeroes}, $0<\nu/p<1$, and that this zero is negative, 
so we thus conclude that 
in this regime the transition occurs at the finite value $b_c =
\sqrt{\frac{\s2 p c}{2}} x_c $, with $x_c<0$ the only zero of $D_{\nu/p}(x)$, 
and is controlled by an exponent $\beta=1/(p-\nu)$, where now $\n \sim (-b+b_c)^\beta$.

We next summarize the scaling regimes obtained for the three cases:

\begin{equation}
\beta =
\begin{cases}
~~~~\frac{\displaystyle 1}{\displaystyle p} \quad ~~ (b_c=0) & {\rm
if} \quad p<\frac{\displaystyle D}{\displaystyle \s2}, \cr \cr
\frac{\displaystyle 1}{\displaystyle \frac{2D}{\s2}-p} \ (b_c=0) & {\rm if} \quad  \frac{\displaystyle D}{\displaystyle \s2}< p< \frac{\displaystyle 2D}{\displaystyle \s2} ,, \cr \cr
\frac{\displaystyle 1}{\displaystyle  p-\frac{2D}{\s2}} \ (b_c<0) & {\rm if} \quad p> \frac{\displaystyle 2D}{\displaystyle \s2}.\cr 
\end{cases}
\label{cases}
\end{equation}

In order to check these results we have solved numerically the
self-consistent equation (\ref{self-cons}). This requires evaluating
numerically the involved integrals.  Figure 1 illustrates the output
of this calculation by showing estimates of $\beta$ as a function of
$p$ and $2D/\s2$. Note the excellent agreement with the analytical
results Eq.(\ref{cases}). In the region around $\s2 p/ 2D \approx 1$,
there are divergences and the integrals are difficult to evaluate
numerically, generating large error bars. For ratios $\s2 p/2D$ larger
than $1$ the location of the critical point obtained by solving
(\ref{self-cons}) numerically is find to coincide with that given by
the zeros of the corresponding parabolic-cylinder function, which we
have also computed numerically. These results provide a complete
verification of the previous analytical calculations.

Let us now consider higher-order moments. We can write for $k\geq 0$
\begin{equation}
m_k\equiv \langle n^k \rangle =\frac{I_k(m)}{I_0(m)}
\end{equation}
with
\begin{equation}
I_k(m)\equiv \int_0^\infty dn \  n^{\gamma+k-1} \exp\bigg\{
-\frac{2b}{\sigma^2 p} n^p-\frac{c}{\sigma^2p}n^{2p}-\frac{\nu m}{n}
\bigg\}=(\nu  m)^{k+\gamma} J^{(k)}
\end{equation}
where 
\begin{equation}
J^{(k)}(m)=\int_0^\infty dx \ x^{\gamma+k-1}
e^{-b^{\prime}(mx)^p-c^{\prime}(mx)^{2p}}e^{-1/x}.
\end{equation}
Splitting the integral into two parts as above , $J^{(k)}=J_1^{(k)}+J_2^{(k)}$,
\begin{equation}
J^{(k)}=J_1^{(k)}+J_2^{(k)}=\int_0^{x_1}dx \ x^{\gamma+k-1} e^{-b'(mx)^p-c'(mx)^{2p}}e^{-\frac{1}{x}}
+\int_{x_1}^\infty dx \ x^{\gamma+k-1} e^{-b'(mx)^p-c'(mx)^{2p}}e^{-\frac{1}{x}}.
\end{equation}
%
%
Proceeding similarly as in the above cases for $J_1$ and $J_2$,  the
lowest powers in $m$ can be identified as 
\begin{equation}
J^{(k)}_1\sim c^{(k)}_1-b'c^{(k)}_2m^p-c^{(k)}_3m^{2p} + ...
\end{equation}
and using (\ref{j2sim})
\begin{equation}
J^{(k)}_2\sim  \hat{c}^{(k)}_1-b'\hat{c}^{(k)}_2m^{-\gamma-k}+...
\end{equation}
Consequently,
\begin{equation}
m_k\sim
\begin{cases} 
m^k \quad       & \text{if } k<\frac{2D}{\s2}+1,\cr
m^{\nu+1} \quad & \text{if } k>\frac{2D}{\s2}+1.\cr
\end{cases}
\end{equation}

This represents a strong form of multiscaling, very similar to the one
reported for the moments in the self-consistent solution for MN1
\cite{colaiori}.

Also, $\langle h \rangle = \langle -\ln n \rangle$ can be computed
effortlessly by making use of
\begin{equation}
\ln n = \lim_{\alpha \to 0} \frac{n^\alpha-1}{\alpha},
\end{equation}
which reduces the calculation of the average of $\ln n$ to a combination of moments of $n$.
Use of this gives $\langle h \rangle \sim -\ln (-b+b_c)$.

Finally, we just mention that the more general situation $bn^{1+p}+cn^{1+q}$ with $q >p$ leads to
the following simple substitutions:
\begin{equation}
\beta =
\begin{cases}
\frac{\displaystyle 1}{\displaystyle q-p} & {\rm if} \quad  p<q<\frac{\displaystyle 2D}{\displaystyle \s2}, \cr
\frac{\displaystyle 1}{\displaystyle \frac{2D}{\s2}-p} & {\rm if} \quad  p< \frac{\displaystyle 2D}{\displaystyle \s2} < q,, \cr
\frac{\displaystyle 1}{\displaystyle p-\frac{\displaystyle 2D}{\displaystyle \s2}} & {\rm if} \quad p> \frac{\displaystyle 2D}{\displaystyle \s2}.\cr
\end{cases}
\end{equation}
To locate the critical point one has now to proceed numerically to search for the zeros of 
$c_5(b)$ which can no longer be expressed in terms of parabolic-cylinder functions.

\subsection{The case MN2}%

Consider again Eq. (\ref{critical_mn}) where for convenience we have introduced
a change in the sign of $p$, which is now positive,
\begin{equation}
\frac{\partial n(x,t)}{\partial t} =D\nabla^2 n +an +bn^{1-p}+cn^{1-2p}
+n\eta(x,t).
\label{critical_mn2}
\end{equation}
This is a non-order-parameter Langevin equation whose associated universality class can be studied 
by measuring the order-parameter $\langle n^{-1} \rangle$. The corresponding stationary
probability density is
\begin{equation}
P_{st}(n) \propto n^{-1+\frac{2(a-D)}{\sigma^2}} \exp
\bigg\{ -\frac{2}{\s2} \bigg( 
\frac{b}{p} n^{-p}+\frac{c}{
2p}n^{-2p}+\frac{Dm}{n}
\bigg)
\bigg\},
\label{pdfmn2}
\end{equation}
with $m =\n \to \infty$ at the transition. Proceeding as in the MN1 case, we define 
\begin{equation}
I(m) =\int_0^\infty dn \  n^{-1+\frac{2(a-D)}{\sigma^2}} \exp \left(
-\frac{2b}{\sigma^2 p} n^{-p}-\frac{c}{\sigma^2p}n^{-2p}-\frac{2D}{\sigma^2}\frac{m}{n}
\right),
\end{equation}
and the self-consistent equation is now given by
\begin{equation}
-\frac{\nu}{m}= \frac{\partial_m I(m)}{I(m)}.
\end{equation}
This last expression can be further simplified by making the change of variable $x=m/n$
and setting as before $a_c=\sigma^2/2$ \cite{mn2} to obtain
\begin{equation}
\frac{\partial_m J(m)}{J(m)}=0,
\label{consistency_mn2}
 \end{equation}
where
\begin{equation}
I(m)=m^{-\nu} J(m), \quad\quad J(m)=\int_0^\infty dx \ x^{\nu-1}
e^{-b^{\prime}(x/m)^p-c^{\prime}(x/m)^{2p}}e^{-\nu/x},
\end{equation}
and $\nu, b',$ and $c'$ are defined as above.
Next, an intermediate point $x_1$ is considered such that $\frac{x_1}{m}<<1$ 
and $J(m)$ is split as in equation (\ref{split}), $J(m)=J_1+J_2$.
The exponential factor $\exp[-b'(x/m)^p-c'(x/m)^{2p}] $ in $J_1(m)$ is expanded, what leads to the 
following asymptotic behavior 
\begin{equation}
J_1(m)\underset{m\sim\infty}{\sim}  c_1-b'c_2m^p-c_3m^{2p}.
\end{equation}
Regarding $J_2$, by virtue of the mean-value theorem
%
\begin{equation}
\frac{dJ_2(m)}{dm}=-b'p m^{-p-1} f_1(\xi)-c'2 p m^{-2p-1}f_2(\xi)
\end{equation}
where $f_i(\xi)$ stands for the function $c_i e^{-b'[\xi_i(m)/m]^p-c^{\prime}[\xi_i(m)/m]^{2p}}$
with $\xi_i(m)\in[x_1,\infty)$ and $c_i=\int_{x_1}^\infty dx \ x^{a_i} e^{-\nu x}$.
Note that $a_i$ does not depend on $m$.
Lastly, after substituting the self-consistent equation reads
\begin{equation}
\frac{d}{dm} \left(c_1-b'c_2 m^{-p}+J_2(m)\right)=0,
\end{equation}
from which it is straightforward to extract the critical point $b_c=0$ and write 
\begin{equation}
m \sim(-b)^{-1/p}.
\end{equation} 

This result has been verified by numerical integration of the
self-consistent equation.

A calculation of the higher moments of the distribution (\ref{pdfmn2})
along the same lines as the previous section (but with $m\to \infty$)
results in $m_k \equiv \langle n^k \rangle \sim m^k$, and therefore
$\nop \sim (-b)^{1/p}$ while, as in MN1, $\langle h \rangle \sim
\ln(m) \sim -\ln(-b)$. Results pertaining to $\nop$ are numerically
verified in Fig. 2.

\section{Discussion and summary}

We have investigated within the mean-field approximation the surface
order-parameter $n=e^{-h}$ at nonequilibrium, critical wetting
transitions of KPZ interfaces interacting with a wall. The model, as
described by the Langevin equation (\ref{critical_mn}), covers both
positive (MN2 $p<0$) and negative (MN1 $p>0$) KPZ nonlinearities.

The more interesting case is that of negative KPZ nonlinearities,
where three different scaling regimes can be distinguished.
\begin{enumerate}

\item 
If $p<D/\s2$ a critical wetting transition exists characterized by the
exponent $\beta=1/p$ and the critical temperature $b_c=0$. We denote
this the {\em pure mean-field regime} because these are the values that
would have obtained had the noise and the Laplacian terms been
neglected in the equation, i.e. in the crudest possible mean-field limit
(\ref{critical_mn}).
\item
If $D/\s2<p<2D/\s2$ the critical wetting temperature is still given by $b_c=0$
but the critical exponent no longer depends solely on the potential details
but also on the noise strength, $\beta=1/(2D/\s2-p)$. This is denoted the {\em weak-noise regime}. 
\item
If $p> 2D/\s2$ the system enters a {\em strong-noise regime} where the
critical temperature is shifted away from zero, $b_c=x_c\sqrt{\s2 pc/2}$
with $x_c<0$ the only zero of the parabolic-cylinder function $D_{2D/\s2 p}(x)$, 
and $\beta=1/(p-2D/\s2)$.
\end{enumerate}

This rich structure is expected to hold at least qualitatively beyond
mean-field theory as it is known that Eq. (\ref{mn}) exhibits a
strong-coupling regime for arbitrary high system dimensionalities. A
similar scenario arises in the full solution of equilibrium critical
wetting with long-ranged forces for which there are also three regimes
for the behavior of $\langle h \rangle$, whose nature is similar to
ours insofar as their origin can be traced back to the relevance of
fluctuations as compared with the potential terms
\cite{reviews_wetting}.  As in the present case, in equilibrium a
first regime exists which is correctly described by naive mean-field
theory.  In the second one the critical temperature is given correctly
by mean-field theory, but the critical exponents are not, and in the
third one fluctuations dominate and mean-field theory has nothing to
say.

An important difference exists, however, between our nonequilibrium
self-consistent solution for negative KPZ nonlinearities and
equilibrium wetting in that the predictions for the former are for
$d=\infty$, while the three-regime behavior for the latter occurs
only below the upper critical dimension $d_c=2$. In higher dimensions
equilibrium wetting is known to be controlled by a Gaussian fixed
point with trivial associated scaling \cite{reviews_wetting}.

Higher-order moments also display interesting behavior. All moments
$m_k$ starting from $2D/\s2+1$ scale with the same exponent, while the
usual scaling $m_k\sim m^k$ obtains for $k\le 2D/\s2+1$. This same
behavior was observed in a mean-field study of nonequilibrium, MN1
complete wetting as reported in \cite{colaiori}, and seems to be a
common feature of multiplicative-noise controlled transitions.

Additionally, it was found that the mean separation at the transition
diverges as $\langle h \rangle \sim -\ln(-b+b_c)$, with $b_c=0$ for
$p<2D/\s2$ and $b_c<0$ for $p>2D/\s2$. This implies that there is a
single scaling regime in terms of $h$ rather than three. That the
behavior of the surface order-parameter is richer than that of the
mean separation is seemingly a common characteristic of these systems.

For positive KPZ non-linearities has also been investigated and
the results show a single regime $\langle n^{-1} \rangle \sim
(-b)^{1/p}$, with higher moments scaling as $m_k \sim m^k$. The mean
interfacial separation grows logarithmically, $\langle h \rangle \sim
\ln m \sim -\ln (-b)$. 

Hence, positive KPZ nonlinearities generate a trivial mean-field (or
high-dimensional) behavior for nonequilibrium critical wetting,
compatible with standard Gaussian scaling, which is analogous to the
mean-field behavior of equilibrium critical wetting. On the contrary,
negative KPZ nonlinearities behave in a rather intricate way, with
highly nontrivial scaling including three different regimes even in
mean-field (or high dimensions) approximation.

In future work we will study how the rich phenomenology reported in
this paper is affected by fluctuations, i.e. going beyond the mean field
approximation. It would also be nice to have nonequilibrium critical
wetting experiments to see whether our predictions can be observed in
real systems.

\section{Acknowledgments}

This work was supported in part by the Spanish projects No. FIS2005-00791 and 
No. FIS2005-00973 (Ministerio de Ciencia y Tecnolog\'ia), FQM-165 and FQM-0207 
(Junta de Andaluc\'ia), and the European project No .INTAS-03-51-6637.

\cleardoublepage
\center{\LARGE{FIGURE CAPTIONS}}

\begin{enumerate}

\item[Fig. 1]
(color online) Estimates of $\frac{2D}{\s2}\beta$ as a function of $\frac{\s2p}{2D}$ 
from power-law fits of $\nop$ vs $b-b_c$ after a numerical integration of Eq. 
(\ref{self-cons}) for MN1. Three regimes are found: the solid (blue) line corresponds to 
$p<D/\s2$, the dashed (red) one to $d/\s2<p<2D/\s2$, and the dash-dotted (green) one to $p>2D/\s2$.
For ratios $2D/\s2<p$, $b_c$ is obtained 
from the (numerically computed) zeros of the corresponding parabolic-cylinder function. 
Close to the singularity $p=D/\s2$ the integration is cumbersome and the resulting errors
large. Far from it, error bars are smaller than the symbols in some cases.

\item[Fig. 2]
(color online) Estimates of $\beta$ as a function of $p$ 
from power-law fits of $\nop$ vs $b-b_c$ after a numerical integration of Eq.
(\ref{self-cons}) for MN2. The error bars are smaller than the symbols.

\end{enumerate}

%
%
%
%
\cleardoublepage


\begin{figure}
\centerline{\psfig{figure=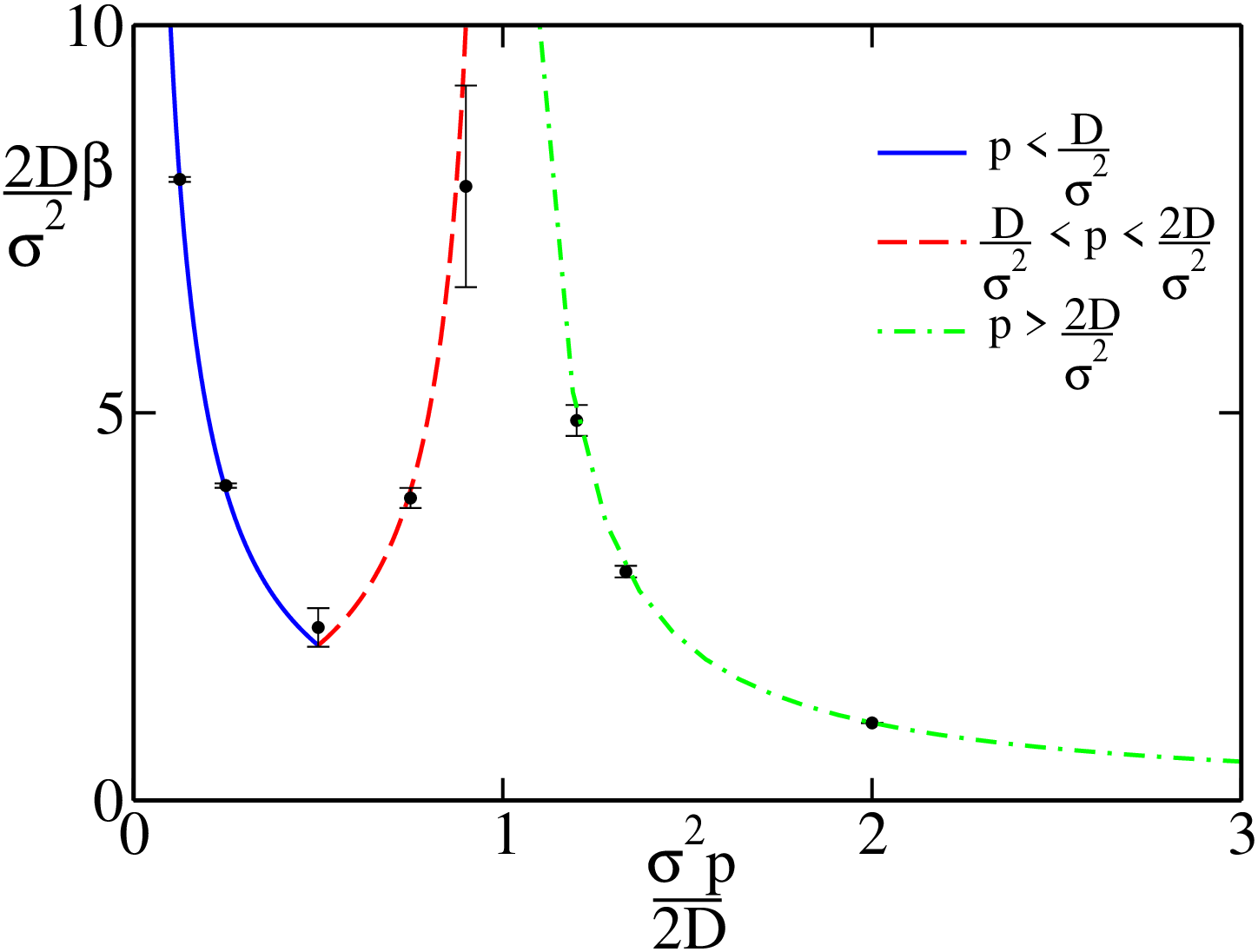,width=12.0cm}}
\caption{}
\label{fig1}
\end{figure}

\cleardoublepage

\begin{figure}
\centerline{\psfig{figure=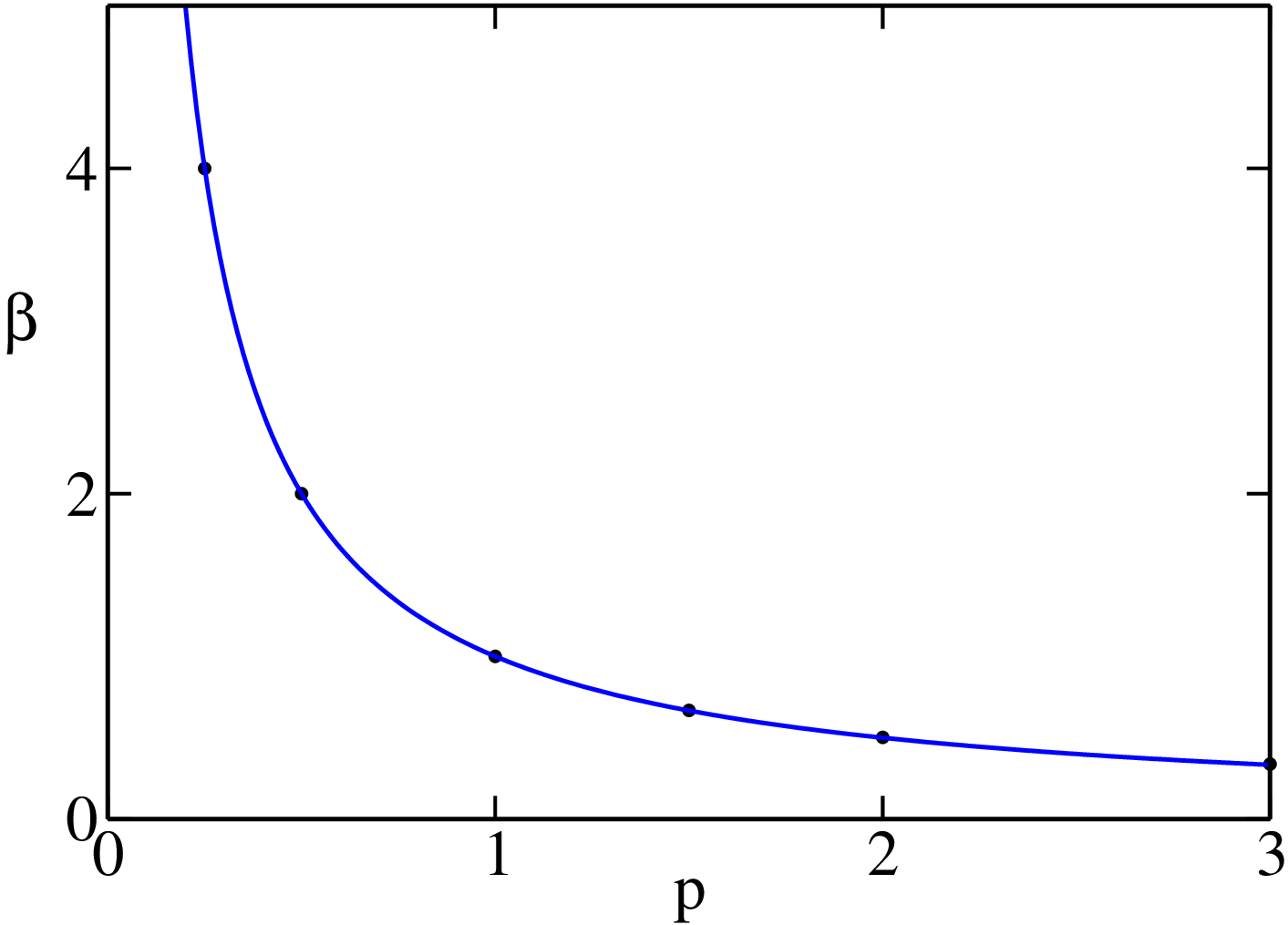,width=12.0cm}}
\caption{}
\label{fig2}
\end{figure}

\end{document}